\documentclass[a4paper,11pt]{article}

\usepackage{pos}
\usepackage{lipsum} 
\usepackage{graphicx} 
\usepackage[utf8]{inputenc}
\usepackage{float}
\usepackage{array}
\usepackage{caption}
\usepackage{subcaption}
\usepackage{wrapfig}
\usepackage{colortbl}
\usepackage{xcolor}

\title{Characterizing IceTop Response to Low-Energy Air Showers}
\ShortTitle{IceTop Response to Low-Energy Air Showers}

\author{The IceCube Collaboration \\{\normalsize \normalfont(a complete list of authors can be found at the end of the proceedings)}\\}

\emailAdd{yanee\_tangjai@cmu.ac.th}
\emailAdd{agnlesz@udel.edu}
\emailAdd{ptatphicha@gmail.com}
\emailAdd{achara.seri@cmu.ac.th}
\emailAdd{tilav@udel.edu}

\abstract{

This study evaluates the response of the IceTop tanks to low-energy air showers in the GeV to TeV energy range based on simulated and measured count rates. Correlating this response with primary cosmic rays provides a tool to study Galactic and solar cosmic-ray flux modulations, particularly for solar particle events. We present long-term behavior of the IceTop scaler rates for a range of discriminator thresholds to better understand and calibrate the detector’s response to changing environmental conditions.

\vspace{4mm}

{\bfseries Corresponding authors:}
Yanee Tangjai$^{1}$, 
Agnieszka Leszczynska$^{3}$, 
Tatphicha Promfu$^{1}$, 
Achara Seripienlert$^{1,2}$, 
Serap Tilav$^{3}$\\
{$^{1}$ \itshape Department of Physics and Materials Science, Faculty of Science, Chiang Mai University, Chiang Mai 50200, Thailand}\\
{$^{2}$ \itshape Office of Research Administration, Chiang Mai University, Chiang Mai 50200, Thailand}\\
{$^{3}$ \itshape Department of Physics and Astronomy, University of Delaware, Newark, DE 19716, USA}\\[4mm]
$^*$ Presenter
}

\ConferenceLogo{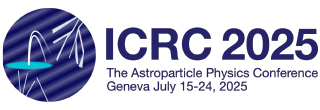}

\FullConference{39th International Cosmic Ray Conference (ICRC2025)\\
 15–24 July 2025\\
Geneva, Switzerland\\}

\begin{document}
\maketitle
\section{Introduction}

The IceTop array\,\cite{IceTop}, which forms the surface component of the IceCube Neutrino Observatory at the South Pole, is comprised of 81 stations, with each station consisting of two ice-Cherenkov tanks spaced 10 meters apart. Each tank contains two Digital Optical Modules (DOMs) with photomultiplier tubes (PMTs), one configured for high gain and the other for low gain to enable a broad dynamic range in signal detection. IceTop is primarily designed for detecting extensive air showers initiated by cosmic rays (CRs) in the energy range of sub-PeV to EeV. In addition to its primary mission, IceTop continuously records the rates of air-shower particles that trigger individual detectors (most of these particles originate from CRs in the GeV-TeV range). These \textit{scaler} rates provide valuable insight not only into the stability of the detector operation, but also into heliospheric phenomena such as solar energetic particles~\cite{SEPson2006} and coronal mass ejections which can cause temporal modulations of Galactic cosmic rays, like Forbush decreases~\cite{Heliospheric}. Moreover, IceTop tanks can observe ground-level enhancements, which are solar-particle events energetic enough to reach the Earth's surface and produce a measurable increase in detector count rates.

IceTop DOMs are equipped with two voltage discriminators. In high gain DOMs, one is used for standard air-shower data taking with a threshold equivalent to about 23 photoelectrons (PEs) collected from the tank. The other is set to lower threshold values, spanning from 2 to 27\,PEs, to configure the whole of IceTop into a multi-channel detector of low-energy cosmic rays.


IceTop scaler rates are affected by environmental factors, namely atmospheric pressure, temperature\,\cite{icecube_atm}, and snow accumulation\,\cite{icetop_snow}. The atmospheric changes influence the development of particle cascades, and hence the distribution of particles reaching the ground. In particular, seasonal changes in temperature influence the atmospheric density, leading to an anti-correlation with the scaler rates. The accumulation of snow on top of the tanks alters the effective overburden, affecting the energy threshold of the particles that reach the detector and the resulting signals observed in the tanks. Without an appropriate correction for these effects, such variations can mimic or obscure the variations originating from the Galactic and solar CR flux. 

This work presents a preliminary analysis of the long-term behavior of IceTop scaler rates spanning from 2013 to 2024 in relation to pressure, temperature, and snow depth. Accurate modeling of these effects, particularly pressure correction, is essential for a prompt response to heliospheric events. To understand the underlying patterns and to correlate the detector response at different discriminator thresholds with the CR energy distributions, we performed Monte Carlo (MC) simulations of a single IceTop tank's response to low-energy air showers. We also studied properties of the air showers generated for atmospheric models of different seasons. 

\section{IceTop scaler rates}

\subsection{Data collection}

We analyze scaler rates collected from May 2013 to December 2024 by all high-gain IceTop DOMs corresponding to various discriminator thresholds. These count rates were recorded at one-minute intervals, facilitating investigations of short-term atmospheric influences as well as long-term CR modulation trends.

Atmospheric surface-pressure data at the South Pole, originally recorded every five minutes, are interpolated to one-minute intervals using cubic spline interpolation from the scipy package\,\cite{scipy} to precisely synchronize with DOM timestamps. The interpolated pressure data support accurate corrections for atmospheric pressure effects and the determination of DOM-specific pressure response coefficients.

To maintain data integrity and eliminate non-physical fluctuations, we apply a consistent outlier rejection procedure across all high-gain DOMs. Specifically, a six-hour running average is calculated from the one-minute count rate data, and points that deviate more than $\pm 5\sigma$ from this local average are removed prior to further analysis.

\begin{figure}
    \centering
     \vspace*{0.5cm}
    \includegraphics[scale=0.28]{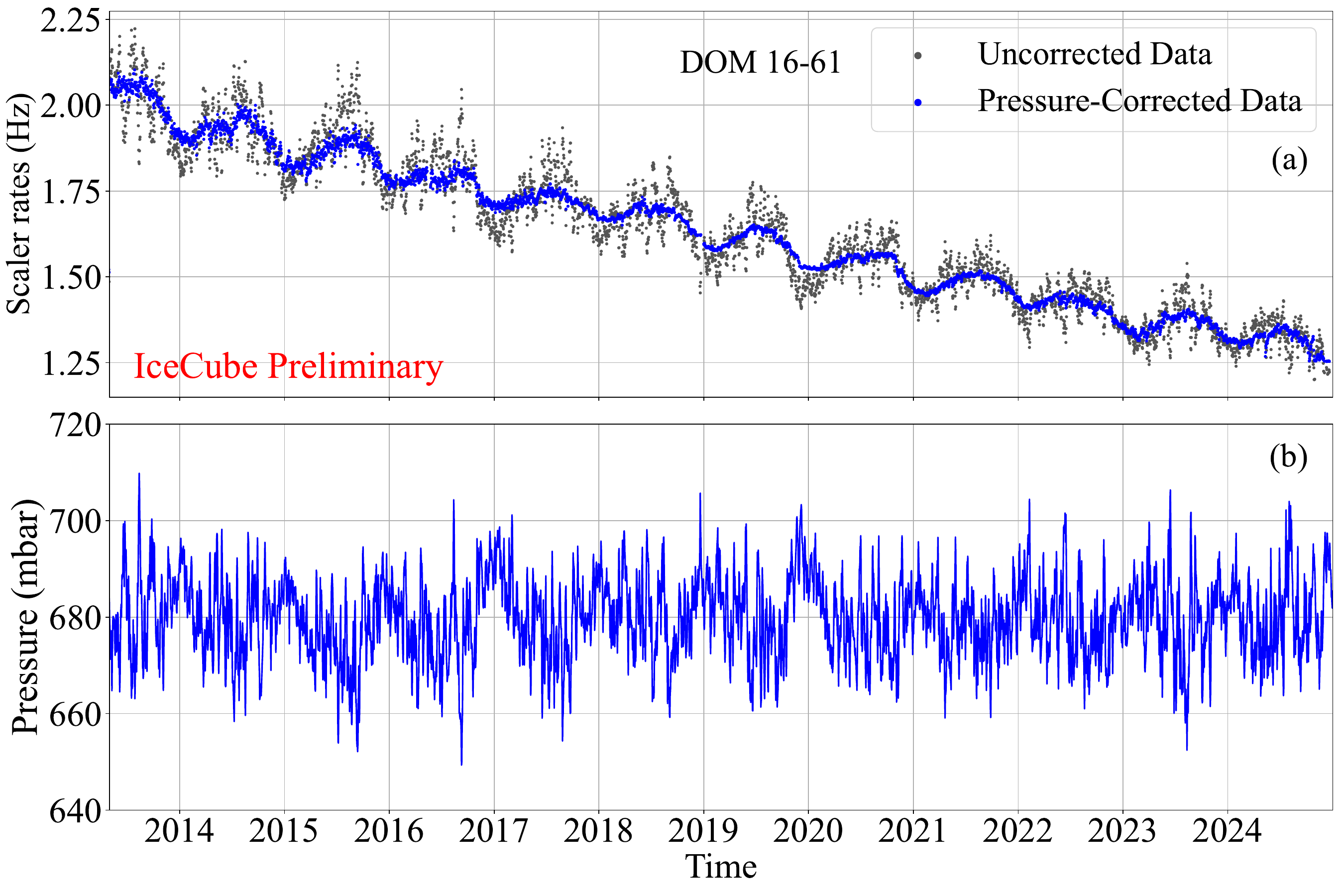}
    \caption{Long-term scaler count rates for IceTop DOM 16–61 from May 2013 to December 2024. Panel (a) shows uncorrected (gray) and pressure-corrected (blue) count rates using daily time bins. Panel (b) displays the corresponding atmospheric pressure in mbar. The pressure correction is applied monthly using the barometric coefficient.}
    \label{fig:longterm_dom1661}
\end{figure}

\subsection{Pressure correction}

Following outlier removal, we quantify the sensitivity of DOM scaler rates to pressure variations by estimating a monthly barometric coefficient, $\beta$, for each DOM. This is done by fitting the relationship between the scaler rate and atmospheric pressure using a linear model in logarithmic space to extract the slope, which corresponds to the barometric coefficient. The resulting $\beta$ values are then applied in a correction model: $C_{\mathrm{corr}} = C_{\mathrm{uncorr}} \cdot \exp[\beta(P - P_0)]$, where $C_{\mathrm{uncor}}$ and $C_{\mathrm{corr}}$ are the uncorrected and corrected scaler rates, $P$ is the atmospheric pressure, and $P_0 = 680\,\mathrm{mbar}$ is the fixed reference pressure determined from the long-term annual average at the South Pole. The fitting is performed independently for each DOM and every month. This correction enables consistent comparison of long-term variations across DOMs and discriminator thresholds.
\begin{figure}
    \centering
     \vspace*{0.5cm}
    \includegraphics[scale=0.29
    ]{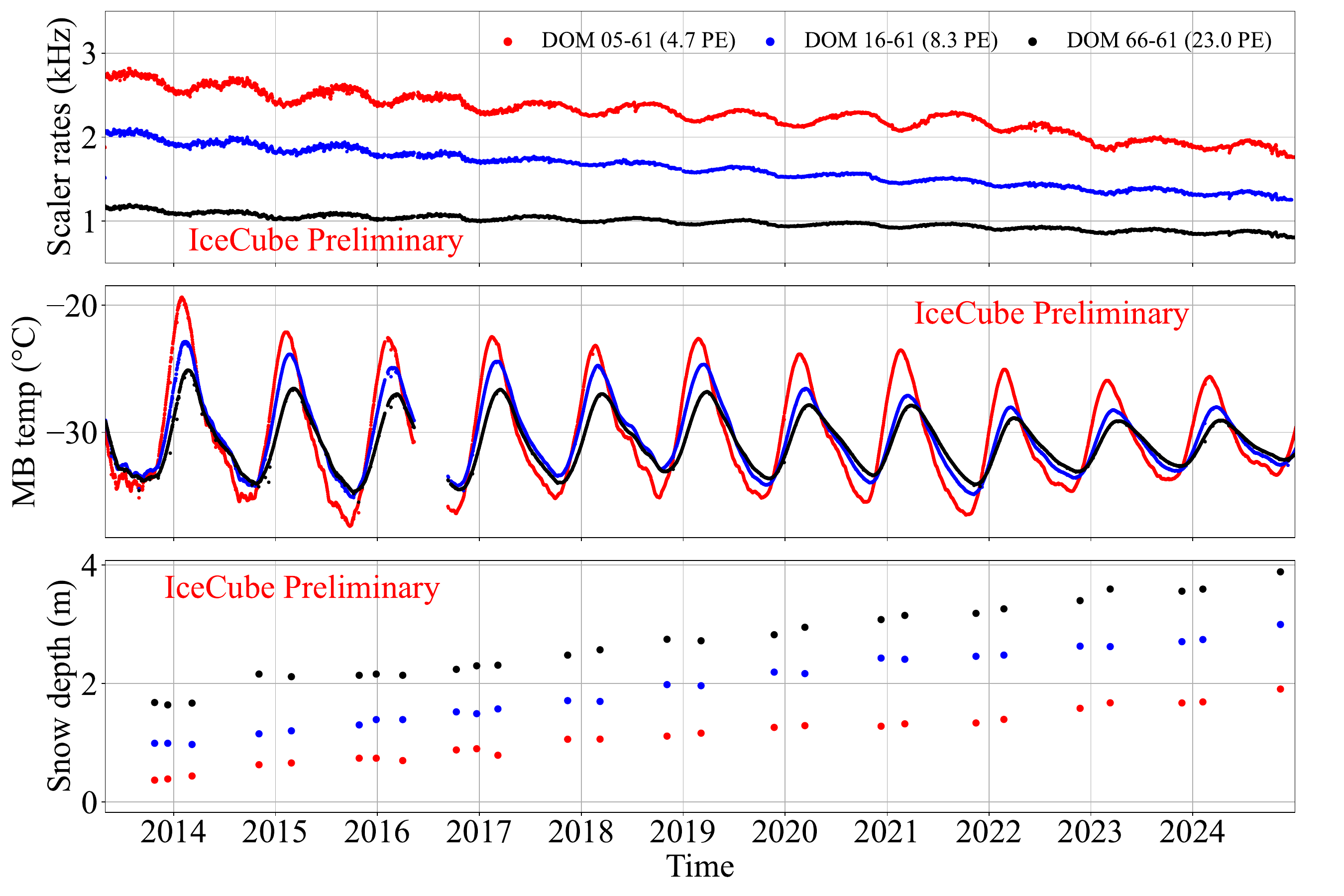}
    \caption{Long-term variations (May 2013 -- December 2024) of IceTop scaler count rates corrected to 680 mbar, mainboard (MB) temperatures, and snow depth. The top panel shows scaler count rates for the following DOMs and their corresponding discriminator thresholds: DOM 05–61, 4.7\,PE (red), DOM 16–61, 8.3\,PE (blue), and DOM 66–61, 23.0\,PE, (black). The middle panel displays daily-averaged mainboard temperatures; the gap in data results from a transition between two data sources. The bottom panel presents in situ snow depth measurements, noting the limited sampling frequency of 2–3 points per year.}
    \label{fig:longterm_mbtemp}
\end{figure}
\subsection{Long-term trend analysis}

To investigate long-term behavior of the IceTop scaler rates, we analyze data from DOM 16–61 as a representative example. Figure~\ref{fig:longterm_dom1661} shows uncorrected (gray) and pressure-corrected (blue) scaler rates from May 2013 to December 2024, along with corresponding atmospheric pressure data. The uncorrected rates exhibit substantial short-term fluctuations that strongly anticorrelate with pressure variations, as expected from the barometric effect. After applying monthly pressure corrections, the corrected rates show significantly reduced variability, making long-term trends more apparent. Data from neutron monitors show that the $\beta$ depends on the incident primary particle spectrum, which is modulated by solar activity~\cite{KOBELEV1}. The long-term behavior of this coefficient for IceTop tanks is the subject of ongoing study.

To further explore environmental effects beyond atmospheric pressure, Figure~\ref{fig:longterm_mbtemp} compares pressure-corrected scaler rates of three DOMs with different discriminator thresholds—DOM 05–61 (4.7\,PE), DOM 16–61 (8.3\,PE), and DOM 66–61 (23.0\,PE)—with their respective mainboard (MB) temperatures and in-situ snow depth measurements. The middle panel shows daily-averaged MB temperatures that exhibit regular annual cycles corresponding to seasonal temperature variations at the South Pole. Previous studies have also shown that scalar rates correlate differently with atmospheric temperatures measured at different altitudes\,\cite{icecube_atm}, which will be the subject of our further research.

Although the scaler rates have been corrected for atmospheric pressure, they still exhibit a gradual long-term decrease, which is inversely correlated with increasing snow depth across the DOMs. For example, DOM 66–61, which has the highest threshold setting (23.0\,PE) and is buried under relatively deep snow, consistently shows the lowest count rate and a weaker apparent seasonal pattern compared to DOMs with lower thresholds. Snow accumulation, however, is not a smooth process, but is a result of wind drift and can therefore change rapidly within a day, leading to a sudden change in the observed rates.
Overall, the lower rates reflect the decreasing detector sensitivity to the secondary particles in CR air showers.

\section{Simulation study}
\subsection{IceTop tank response}
\begin{wrapfigure}{r}{0.55\textwidth}
    \centering 
     \includegraphics[width=0.99\linewidth]{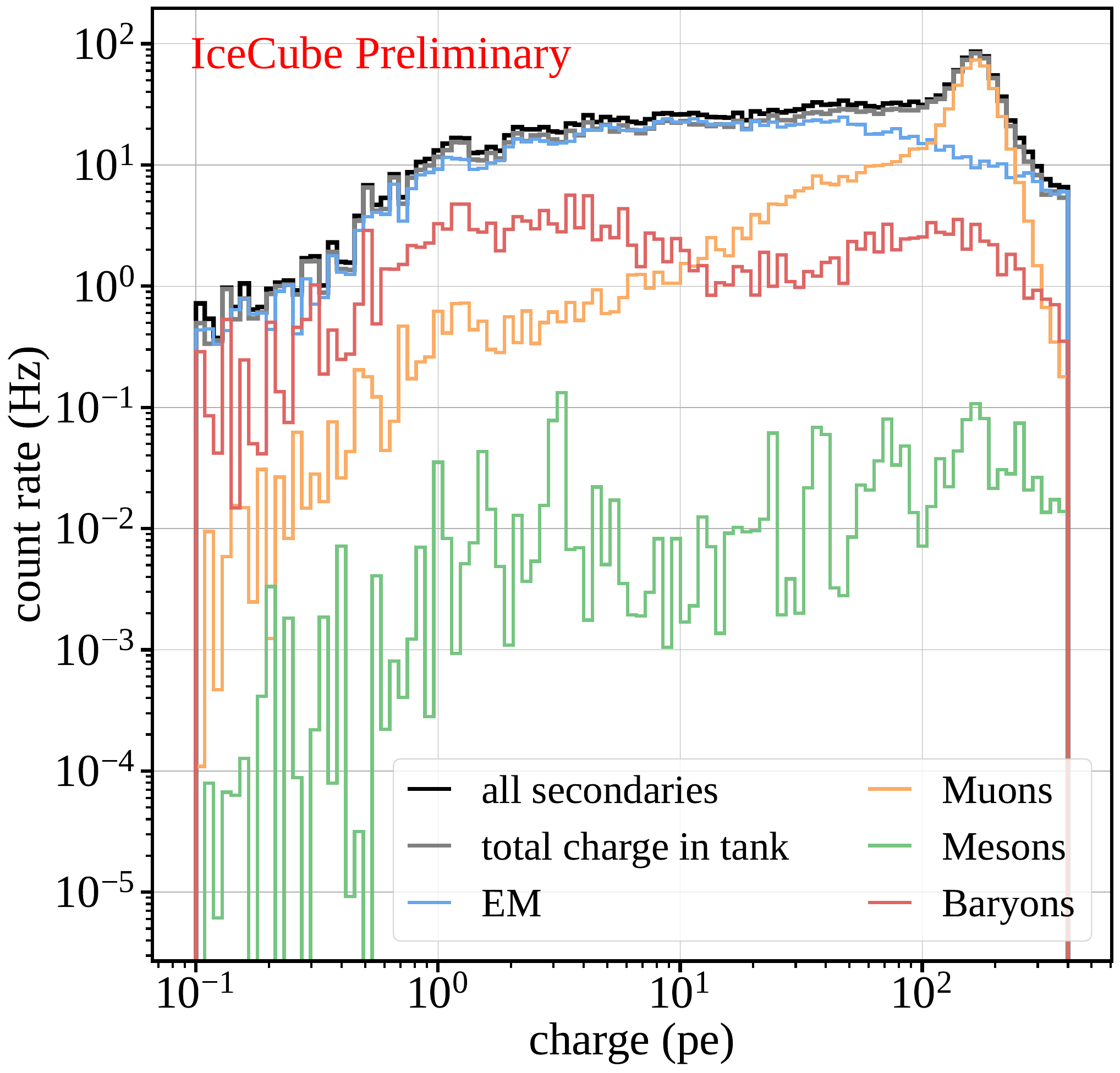}
  \caption{Charge spectrum observed by a high-gain DOM of the IceTop tank (buried under 2.46\,m snow) in response to secondary particles from 1\,GeV -- 125\,TeV primary protons, where each entry corresponds to a single particle deposition. In addition, the gray line indicates the spectrum of total charge in the tank - higher due to presence of multiple particles contributing to the charge.
}
    \label{fig:weighted_charge}
\end{wrapfigure}
In order to understand the observed patterns and correlate the scaler rates with the primary CR spectrum, we conduct preliminary MC studies.
The air-shower simulations were performed using the CORSIKA program~\cite{corsika} with FLUKA~\cite{fluka1, fluka2} as low-energy hadronic interaction model and Sibyll 2.3d~\cite{sibyll} for high-energy interactions.
The secondary particles were simulated above a certain kinetic energy threshold, which has been set to 50\,MeV for hadrons (without $\pi^0$), 10\,MeV for muons and electrons, and 2\,MeV for photons (including $\pi^0$).
We simulated 991,800 proton primaries over an energy range from 1\,GeV to 125\,TeV. The generated spectrum follows approximately $E^{-2}$ up to 300\,GeV and $E^{-1}$ above. The zenith angle $\theta$ of the incoming primaries is distributed between $0^\circ$ and $85^\circ$, assuming equal intensity with the solid angle and taking into account geometrical efficiency of a horizontal detector. CRs with energy below 3\,GeV did not generate any signals in the simulated tank in our MC sample. 

Air-shower particles undergo multiple scattering, and some of them reach the ground kilometers away from the air-shower axis. To account for them in the tank response, the air showers are resampled so that trajectory of each (or at least some fraction) of these particles can geometrically cross the chosen detector, resulting in a different number of resamples for each simulated air shower. Therefore, each event $i$ is given the following weight:
\begin{equation}
w(E_i, N_k, A_i) = \frac{A_i \, \Omega \, J_{\mathrm{CR}}(E_i)}{N_k \, J_{\mathrm{MC}}(E_i)} \int_{E_k}^{E_{k+1}} J_{\mathrm{MC}}(E) \, dE
\label{eq:weights}
\end{equation}
where $A_i$ is the effective area over which this particular air shower can be thrown to have a chance of generating any signal in the chosen tank, $\Omega$ is simulated solid angle ($0^\circ$ - $85^\circ$), $J_{\mathrm{MC}}(E_i)$ is the simulated flux, $J_{\mathrm{CR}}(E_i)$ is the expected CR flux at $E_i$, and $N_k$ is the number of simulated re-samplings in energy bin $k$. For the expected CR flux, the all-particle energy spectrum of the H4a model was used~\cite{H4amodel}, assuming all CRs are protons.

The IceTop tank response was simulated using Geant4~\cite{geant4} embedded in the IceCube software framework. The results are evaluated for the high-gain DOM of tank 16A, which is buried under 2.46\,m of snow as measured in November 2021. Figure~\ref{fig:weighted_charge} shows the charge spectrum (weighted according to equation\,\ref{eq:weights}) in response to the simulated proton flux. The obtained charge represents the photoelectrons produced by PMT (pe) of that DOM before the discriminator simulation. The counts are separated into secondary-particle groups. While the rates are dominated by electromagnetic (EM) particles at low PE levels, there is also a significant contribution from baryons, mainly protons and neutrons. Moreover, a clear through-going muon peak is visible, which is used to calibrate the detectors. Mesons, mainly pions, contribute the least to the rate. 

\subsection{Air showers under different atmospheric conditions}
\begin{figure}
    \centering
    \includegraphics[scale=0.27]{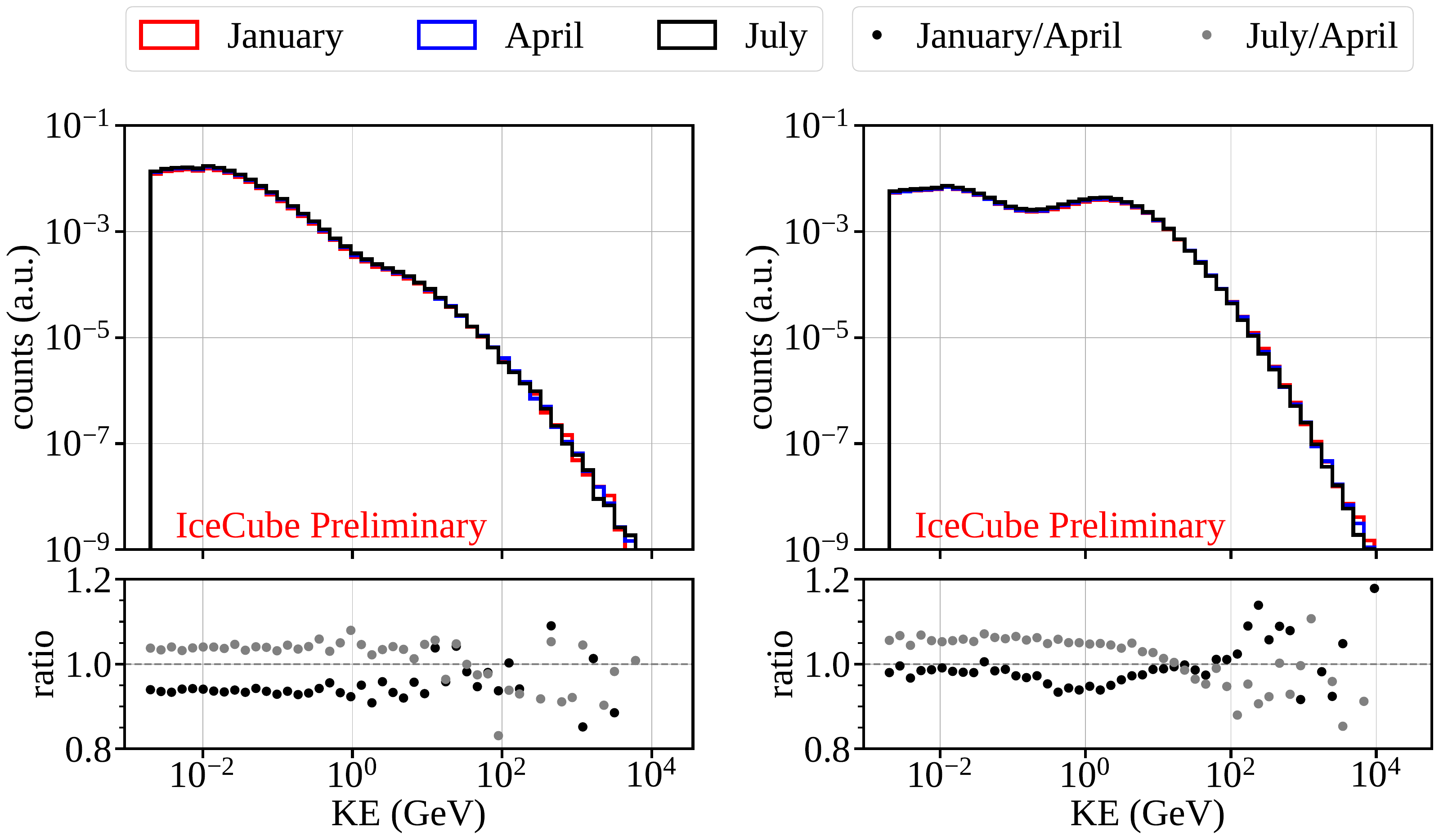}
    \caption{Ground-level kinetic energy spectra of muons and EM particles originating from baryon grandmother, under different seasonal atmospheric models. The left panel shows particles coming predominantly from $\pi^{0}$'s, while the right plot shows mainly products of $\pi^{\pm}$'s. See text for more details.}
    \label{fig:ehistory_all}
\end{figure}

The air-shower particles differ in experiencing the influence of the changes in the atmosphere. These differences do not depend only on the type of particles, but also on the type of particles from which they originate. To study that, we simulated low-energy air showers for three South Pole atmospheric profiles, corresponding to January, April, and July. We use data generated solely from CORSIKA simulations (without detector response) with the EHISTORY option enabled, which records the information about mothers and grandmothers of muons and EM particles. We consider here only the ground-level particles whose grandmother particle is a baryon, which we assume is the tracer of the primary proton.
We further separate the ground level particles into those which contain a $\pi^{\pm}$ decay in their history, and those that do not - hence dominated by $\pi^0$ decay also including $K$'s. The distributions of kinetic energy of ground particles for these two groups are presented in Figure\,\ref{fig:ehistory_all}. We note a dip in the number of $\sim$~GeV particles with $\pi^{\pm}$ decay for the January atmosphere, representing the austral summer. In January, the muon production height is increased due to the warmer temperature, and it is less likely for GeV muons to survive to the ground. Muons from $K$ decay, in the left panel, are not as prevalent and this feature is not prominent. The seasonal variation in the left panel is plausibly explained due to a {15 mbar} difference in pressure between the January and July model atmospheres affecting the EM cascades following $\pi^0$ decay. 
%
%
%
Further studies, including particles from non-baryon grandmothers, are ongoing. Studying the contributions of different particle groups to the overall scaler rates is relevant for accurate modeling of the environmental corrections, given their differing response to the atmospheric modulations.

\section{Summary}
IceTop scaler rates can be used to observe modulations of the Galactic and solar CR flux originating from solar activity. The changing atmosphere requires normalizing these rate results so that only the effects associated with solar activity remain. The conditions at the South Pole bring an additional challenge in the form of snow accumulation, which affects the response of the detectors, and thus the relationship with the measured CR spectrum. This work continues the previous efforts~\cite{groundEnhanc} to establish a model based on collected data and a detailed MC study in which all these effects can be accounted for, allowing observation of transient as well as long-term CR flux modulations originating from solar activity.

Further work will focus on optimizing the pressure correction procedure and modeling the effect of other environmental factors, such as snow accumulation, temperature of the motherboard and also of the atmosphere\,\cite{icecube_atm}. We will extend our simulations to account for neutron production deep in the atmosphere, followed by thermalization and absorption in the tanks or nearby snow accumulation. In addition, it will be necessary to consider relevant background sources and systematic factors affecting scaler rates, especially at the lowest discriminator thresholds.



\begin{thebibliography}{10}

\bibitem{IceTop}
{\bfseries IceCube} Collaboration, \href{http://dx.doi.org/https://doi.org/10.1016/j.nima.2012.10.067}{{\em Nuclear Instruments and Methods in Physics Research Section A: Accelerators, Spectrometers, Detectors and Associated Equipment} {\bfseries 700} (2013) 188--220}.

\bibitem{SEPson2006}
{\bfseries IceCube} Collaboration, \href{http://dx.doi.org/10.1086/595679}{{\em The Astrophysical Journal} {\bfseries 689} no.~1, (2008) L65–L68}.

\bibitem{Heliospheric}
{\bfseries IceCube} Collaboration, T.~Kuwabara, J.~W. Bieber, and R.~Pyle, ``\href{https://indico.nucleares.unam.mx/event/4/session/8/contribution/729/material/paper/0.pdf}{Heliospheric physics with {I}ce{T}op},'' in {\em Proc. of the 30th ICRC (ICRC2007)}, vol.~1, pp.~339--342.
\newblock 2008.

\bibitem{icecube_atm}
{\bfseries IceCube} Collaboration, \href{http://dx.doi.org/https://doi.org/10.48550/arXiv.1001.0776}{``Atmospheric variations as observed by {I}ce{C}ube,''} in {\em Proc. of the 31st ICRC}.
\newblock 2009.

\bibitem{icetop_snow}
{\bfseries {IceCube}} Collaboration, ``\href{https://inspirehep.net/files/e376d37bfb70ca08a10cb1387418472f}{The Effect of Snow Accumulation on Signals in {I}ce{Top}},'' in {\em {Proc. of the 33rd ICRC}}, p.~1106.
\newblock 2013.

\bibitem{scipy}
P.~Virtanen {\em et~al.}, \href{http://dx.doi.org/10.1038/s41592-019-0686-2}{{\em Nature Methods} {\bfseries 17} (2020) 261--272}.

\bibitem{KOBELEV1}
P.~Kobelev, A.~Belov, E.~Mavromichalaki, M.~Gerontidou, and V.~Yanke, ``\href{https://galprop.stanford.edu/elibrary/icrc/2011/papers/SH4.2/icrc0654.pdf}{Variations of Barometric Coefficients of the Neutron Component in the 22-23 Cycles of Solar Activity},'' in {\em Proc. of the 32nd ICRC (ICRC2011)}, vol.~11, pp.~382--385.
\newblock 2011.

\bibitem{corsika}
D.~Heck {\em et~al.}, \href{http://dx.doi.org/10.5445/IR/270043064}{``{CORSIKA}: A {M}onte {C}arlo code to simulate extensive air showers,''} tech. rep., 1998.
\newblock 51.02.03; LK 01; Wissenschaftliche Berichte, FZKA-6019 (Februar 98).

\bibitem{fluka1}
A.~Ferrari, P.~R. Sala, A.~Fasso, and J.~Ranft. \href{https://doi.org/10.5170/CERN-2005-010}{CERN-2005-10, SLAC-R-773 (2005)}.

\bibitem{fluka2}
T.~T. B{\"o}hlen {\em et~al.}, \href{http://dx.doi.org/10.1016/j.nds.2014.07.049}{{\em Nucl. Data Sheets} {\bfseries 120} (2014) 211--214}.

\bibitem{sibyll}
F.~Riehn, R.~Engel, A.~Fedynitch, T.~K. Gaisser, and T.~Stanev, \href{http://dx.doi.org/10.1103/PhysRevD.102.063002}{{\em Phys. Rev. D} {\bfseries 102} (Sep, 2020) 063002}.

\bibitem{H4amodel}
T.~K. Gaisser, \href{http://dx.doi.org/https://doi.org/10.1016/j.astropartphys.2012.02.010}{{\em Astroparticle Physics} {\bfseries 35} no.~12, (2012) 801--806}.

\bibitem{geant4}
{\bfseries GEANT4} Collaboration, S.~Agostinelli {\em et~al.}, \href{http://dx.doi.org/10.1016/S0168-9002(03)01368-8}{{\em Nucl. Instrum. Meth.} {\bfseries A506} (2003) 250--303}.

\bibitem{groundEnhanc}
{\bfseries IceCube} Collaboration, T.~Kuwabara and P.~Evenson, ``Ground level enhancement of {M}ay 17, 2012 observed at {S}outh {P}ole,'' in {\em Proc. of the 33rd ICRC (ICRC2013)}, vol.~33, p.~1347.
\newblock 2013.

\end{thebibliography}
\providecommand{\href}[2]{#2}\begingroup\raggedright\endgroup

%

\clearpage

\section*{Full Author List: IceCube Collaboration}

\scriptsize
\noindent
R. Abbasi$^{16}$,
M. Ackermann$^{63}$,
J. Adams$^{17}$,
S. K. Agarwalla$^{39,\: {\rm a}}$,
J. A. Aguilar$^{10}$,
M. Ahlers$^{21}$,
J.M. Alameddine$^{22}$,
S. Ali$^{35}$,
N. M. Amin$^{43}$,
K. Andeen$^{41}$
% ...
\\
\\
% ...
$^{10}$ Universit{\'e} Libre de Bruxelles, Science Faculty CP230, B-1050 Brussels, Belgium \\
% ...
$^{16}$ Department of Physics, Loyola University Chicago, Chicago, IL 60660, USA \\
$^{17}$ Dept. of Physics and Astronomy, University of Canterbury, Private Bag 4800, Christchurch, New Zealand \\
% ...
$^{21}$ Niels Bohr Institute, University of Copenhagen, DK-2100 Copenhagen, Denmark \\
$^{22}$ Dept. of Physics, TU Dortmund University, D-44221 Dortmund, Germany \\
% ...
$^{35}$ Dept. of Physics and Astronomy, University of Kansas, Lawrence, KS 66045, USA \\
% ...
$^{39}$ Dept. of Physics and Wisconsin IceCube Particle Astrophysics Center, University of Wisconsin{\textemdash}Madison, Madison, WI 53706, USA \\
% ...
$^{41}$ Department of Physics, Marquette University, Milwaukee, WI 53201, USA \\
% ...
$^{43}$ Bartol Research Institute and Dept. of Physics and Astronomy, University of Delaware, Newark, DE 19716, USA \\
% ...
$^{63}$ Deutsches Elektronen-Synchrotron DESY, Platanenallee 6, D-15738 Zeuthen, Germany \\
$^{\rm a}$ also at Institute of Physics, Sachivalaya Marg, Sainik School Post, Bhubaneswar 751005, India \\
% ...

\subsection*{Acknowledgments}

\noindent
The authors gratefully acknowledge the support from the following agencies and institutions:
USA {\textendash} U.S. National Science Foundation-Office of Polar Programs,
U.S. National Science Foundation-Physics Division,
U.S. National Science Foundation-EPSCoR,
U.S. National Science Foundation-Office of Advanced Cyberinfrastructure,
Wisconsin Alumni Research Foundation,
Center for High Throughput Computing (CHTC) at the University of Wisconsin{\textendash}Madison,
Open Science Grid (OSG),
Partnership to Advance Throughput Computing (PATh),
Advanced Cyberinfrastructure Coordination Ecosystem: Services {\&} Support (ACCESS),
Frontera and Ranch computing project at the Texas Advanced Computing Center,
U.S. Department of Energy-National Energy Research Scientific Computing Center,
Particle astrophysics research computing center at the University of Maryland,
Institute for Cyber-Enabled Research at Michigan State University,
Astroparticle physics computational facility at Marquette University,
NVIDIA Corporation,
and Google Cloud Platform;
Belgium {\textendash} Funds for Scientific Research (FRS-FNRS and FWO),
FWO Odysseus and Big Science programmes,
and Belgian Federal Science Policy Office (Belspo);
Germany {\textendash} Bundesministerium f{\"u}r Forschung, Technologie und Raumfahrt (BMFTR),
Deutsche Forschungsgemeinschaft (DFG),
Helmholtz Alliance for Astroparticle Physics (HAP),
Initiative and Networking Fund of the Helmholtz Association,
Deutsches Elektronen Synchrotron (DESY),
and High Performance Computing cluster of the RWTH Aachen;
Sweden {\textendash} Swedish Research Council,
Swedish Polar Research Secretariat,
Swedish National Infrastructure for Computing (SNIC),
and Knut and Alice Wallenberg Foundation;
European Union {\textendash} EGI Advanced Computing for research;
Australia {\textendash} Australian Research Council;
Canada {\textendash} Natural Sciences and Engineering Research Council of Canada,
Calcul Qu{\'e}bec, Compute Ontario, Canada Foundation for Innovation, WestGrid, and Digital Research Alliance of Canada;
Denmark {\textendash} Villum Fonden, Carlsberg Foundation, and European Commission;
New Zealand {\textendash} Marsden Fund;
Japan {\textendash} Japan Society for Promotion of Science (JSPS)
and Institute for Global Prominent Research (IGPR) of Chiba University;
Korea {\textendash} National Research Foundation of Korea (NRF);
Switzerland {\textendash} Swiss National Science Foundation (SNSF).

\end{document}